\documentstyle[preprint,aps]{revtex}
\begin{document}
% \draft command makes pacs numbers print
\draft
\title{Interatomic Potentials from First-Principles Calculations:\\
the Force-Matching Method}
% repeat the \author\address pair as needed
\author{Furio Ercolessi}
\address{
Materials Research Laboratory, 
University of Illinois,
104 S. Goodwin Ave., Urbana, IL 61801, USA \\
International School for Advanced Studies (SISSA-ISAS), 
Via Beirut 4, I-34014 Trieste, Italy
}
\author{James B. Adams}
\address{
Dept.\ of Materials Science and Engineering,
University of Illinois,
105 S. Goodwin Ave., Urbana, IL 61801, USA
}

\date{June 26, 1993}
\maketitle
\begin{abstract}
We present a new scheme
to extract numerically ``optimal'' interatomic potentials
from large amounts of data produced by first-principles calculations.
The method is based on fitting the potential to {\em ab initio}
atomic forces of many atomic configurations, 
including surfaces, clusters, liquids and crystals at finite temperature.
The extensive data set overcomes
the difficulties encountered by traditional fitting approaches
when using rich and complex analytic forms, 
allowing to construct potentials with 
a degree of accuracy comparable to that obtained
by {\em ab initio} methods.
A glue potential for aluminum obtained with this method
is presented and discussed.
\end{abstract}


%\narrowtext
While first-principles methods for computer simulation 
in condensed matter are rapidly improving in speed and accuracy,
classical interatomic potentials continue to constitute
the only way to perform molecular dynamics (MD) or Monte Carlo
computations on systems with a very large size 
(number of atoms $N\sim 10^4$--$10^6$)
or for long simulation times ($t\sim$ nanoseconds).
With the advent of massively parallel machines and the
proper computer codes,
simulations on the mesoscopic scale appear feasible,
allowing one to address a whole new range of
problems in the physics of defects, surfaces, clusters, liquids
and glasses.
However, obtaining accurate and realistic potentials
constitutes a challenging problem.
It is now well recognized \cite{carlsson} that
fairly elaborate analytic expressions---involving for instance
density-dependent terms, angular forces, or moment expansions---are
necessary for a realistic description of most materials under 
different conditions (geometries, structures, thermodynamic phases).
A typical potential is thus constituted by a number of
functions combined in a complex way, and often nested one into another.
Unfortunately, such powerful forms can make the task of fitting
a potential to a given material quite formidable and cumbersome.
There are often many possible ways to fit a set of
experimental quantities within a given analytic framework, and
rather arbitrary assumptions on the functions are usually made 
to reduce the number of parameters to a manageable level.
Such assumptions could be the basic reason why 
potentials apparently good at $T=0$ sometimes fail at finite temperature,
or for geometries or local conditions not considered when the fit was made.

On the other hand, the development of first-principles
methods---where forces on atoms are obtained by directly solving
the electronic structure problem---has
been very vigorous in the last decade
\cite{carpar,sankey}, and evolved to a point
where dynamical simulations
of systems with $N$ of the order of 100--1000
and times $t$ of the order of picoseconds
are within reach for an ever increasing number of physical systems.
Therefore, it seems compelling to
construct a bridge between these two research lines,
making use of the large amount of information 
that can be obtained by first-principles methods
to construct reliable potentials for computations
on a much larger scale.

While one possible way to achieve this consists of trying
to derive potentials from first-principles theoretically
by exploiting approximation schemes \cite{jacobsen,chetty},
we proceed here along different, somewhat complementary lines.
Namely, with realism of the final potential as the main goal,
we present a new method 
to process a large amount of output of first-principles
calculations (positions and forces), and combine this information
with traditional fitting on experimental quantities,
obtaining a potential by a numerical optimization procedure.

We assume that the analytic form of the potential
is defined by a number of single variable functions whose
arguments are simply obtained from the atomic coordinates.
Such is the case for essentially all the two-body and many-body
potentials currently in use.
For example, a glue potential \cite{glue,eam,finnis}
\begin{equation}
V={1\over 2}\sum_{ij}\phi(r_{ij}) + \sum_i U\biggl(\sum_j \rho(r_{ij})\biggr)
\label{eq:glue}
\end{equation}
is defined by
a pair potential $\phi(r)$, an ``atomic density'' function $\rho(r)$,
and a ``glue function'' $U(n)$.
Other potentials may have functions of bond angles or
of other quantities.
Let $\alpha$ indicate the entire set of $L$ parameters 
$\alpha_1, \ldots ,\alpha_L$ used to characterize the functions.
To determine the ``optimal'' set $\alpha^*$
we try to match the forces supplied by
first-principles calculations for a large set of different 
configurations with those predicted by the classical potential,
by minimizing the objective function
\begin{equation}
Z(\alpha) = Z_F(\alpha) + Z_C(\alpha)
\label{eq:z}
\end{equation}
with
\begin{equation}
Z_F(\alpha) =
\biggl(3\sum_{k=1}^M N_k\biggr)^{-1}
\sum_{k=1}^M \sum_{i=1}^{N_k}
  \left|
     {\bf F}_{ki} (\alpha) - {\bf F}_{ki}^\circ
  \right|^2 ,   \label{eq:zf}
\end{equation}
\begin{equation}
Z_C(\alpha) =
  \sum_{r=1}^{N_C} W_r [ A_r(\alpha) - A_r^\circ ]^2 . \label{eq:zc}
\end{equation}
In $Z_F$, $M$ is the number of sets of atomic configurations available,
$N_k$ is the number of atoms present in configuration $k$,
${\bf F}_{ki} (\alpha)$ is the force on the $i$-th atom in
set $k$ as obtained with parametrization $\alpha$,
and ${\bf F}_{ki}^\circ$ is the reference force from first-principles.
$Z_C$ contains contributions from $N_C$ additional constraints.
$A_r (\alpha)$ are physical quantities as calculated with
parametrization $\alpha$.
$A_r^\circ$ are the corresponding reference quantities, which may be supplied
either from the first principles calculation, or perhaps
more likely directly from experimental data.
$W_r$ are weights which are chosen at convenience.

%Forces have been chosen instead of e.g.\ energies,
%since they are available for each atom in each
%configuration and therefore contain a large amount of
%information on the potential energy surface for a given
%local environment.
The $M$ configurations do not need
to be related to each other, and in fact it is
desirable to include input data
relative to different geometries and physical situations
(clusters, surfaces, bulk, solid, defects, liquid, etc.)
in the attempt to achieve a good potential transferability.
In practice, one can use samples from
high temperature {\em ab initio\/} MD trajectories
for various systems, thus obtaining a good representation
of the regions of configuration space that they actually explore
at finite $T$.

Expression (\ref{eq:z}) can be seen under two different points
of view.
If emphasis is given to $Z_F$,
then the minimizing potential appears as the ``best'' approximation
of the first-principles system, with $Z_C$
acting as a guide towards 
the correct region in the space of parameters.
In fact, many properties (for instance, the formation energy of a 
defect) are not easily determined from the forces alone.
Moreover, one could include terms aimed at correcting
shortcomings of the first-principles method in use.
If emphasis is given, instead, to $Z_C$,
then the method looks like a conventional fit,
but where the $Z_F$ term relieves the researcher
from the enormous burden of guessing
the shape of the functions constituting the model.
$A_r^\circ$ and $W_r$ can be maneuvered for tuning.

It should also be noted that invariance properties of the Hamiltonian
must be recognized and taken care of by imposing
additional, dummy constraints.
For instance, a glue potential (\ref{eq:glue}) is invariant under the
transformations 
(a) $\rho(r)\rightarrow A\rho(r) , U(n)\rightarrow U(n/A)$, and
(b) $\phi(r)\rightarrow\phi(r)+2B\rho(r) , U(n)\rightarrow U(n)-Bn$
\cite{glue}.
The two constants $A$ and $B$ are arbitrary and must be fixed
by additional conditions,
%In the application presented below,
%we have chosen $n_{\rm bulk} = 1$ and $U'(n_{\rm bulk})=0$,
%where $n_{\rm bulk}=\sum_j \rho(r_{ij})$ around an atom $i$
%in a perfect fcc crystal at equilibrium.
that can be enforced
as further quadratic terms in eq.\ (\ref{eq:zc}), 
as if they were constraints for physical properties.
In contrast with the latter,
these terms exactly vanish at the minimum.

In the present realization, the single variable functions constituting
the potential are defined as third-order polynomials (cubic splines)
connecting a set of points $\beta_l$, preserving continuity
of the functions and of their first two derivatives across the junctions.
The parameters $\alpha_l$ are a one-to-one mapping to the points
$\beta_l$, chosen on the basis of computational convenience
\cite{boston,long}.  In the simplest case, $\beta_l = \alpha_l$.
%While the points $\beta$ could be directly used as parameters,
%we have found that convergence is faster when the actual parameters
%are defined as differences $\alpha_l = \beta_{l+1}-\beta_l$.
%This is because forces
%are more directly connected with the first derivatives
%rather than with the functions themselves.
Particular boundary conditions, such as requiring a function
and its first derivative to be zero
at a cutoff distance $R_c$, are directly incorporated into
the parametrization.
A number of parameters of the order of 10--20 per function seems
to give sufficient flexibility, and in fact using finer grids
may give rise to noise problems (oscillations of the functions 
depending on the input set) in the spline-based formulation.
%The total Hamiltonian, depending on its complexity, can
%therefore be characterized by a number of parameters $L$
%in the 50--100 range.

The computational engine of the method is a multidimensional
minimization procedure for the objective function (\ref{eq:z}).
To be prepared to deal with the presence of multiple local
minima, we have implemented a simulated annealing algorithm
in parameter space (described in \cite{boston,long}).
However, we found it to be necessary only when starting
from an initial guess very far from the optimal one.
The basin of attraction of the optimal potential is sufficiently
broad that a simple quasi-Newton method is adequate
to reminimize $Z$ after small adjustments to the
values of $A_r^\circ$ and $W_r$, or changes in the set
of first-principles configurations.
In a typical minimization run, $Z$ is evaluated a few thousand
times.
The force computations are carried out by using standard
MD techniques such as neighbor lists to decrease computer time,
and the computational resources can be compared
to those requested by a classical MD code.
In preliminary tests using MD trajectories generated by
classical potentials \cite{boston},
this scheme has proven to be able
to reconstruct {\em exactly} the original potentials
without any further assumption beyond the analytic form---within
the precision allowed by the spacing between spline knots,
and within the range of the function arguments sampled by the input data.

The first application of the force-matching method presented
here is for aluminum.  The reasons for this choice are threefold:
{\em (i)} due to the absence of $d$ electrons, Al can be studied
quite easily and accurately with present first-principles methods;
{\em (ii)} the metallic character suggests the use of
the relatively simple glue model (\ref{eq:glue}), even
if it has well known limitations (lack of angular forces);
{\em (iii)} in spite of the simpler electronic structure,
glue-like potentials obtained so far for Al seem to be less accurate
than those for noble metals, and the validity of glue schemes for
Al has been recently questioned \cite{robertson}, 
making it worthwhile to investigate this issue further.

The potential has been parametrized by a total of $L=40$
parameters (spline knots), of which 14 for $\phi(r)$ and
$\rho(r)$, and 12 for $U(n)$.
The functions reach zero at $r=R_c$, and at $n=0$,
by means of fixed additional spline knots.
The first-principles data used as input have been extracted by 
trajectories of MD simulations using the
local orbital density functional scheme described in \cite{sankey}.
We have processed a total of $M=85$ sets of atomic configurations,
of which 7 represent a 
bulk system with a vacancy ($N=107$) at $T=100$ K,
10 the same system at 1750 K (undergoing melting), 
20 an equilibrated bulk liquid ($N=108$) at 2650 K,
13 a (100) slab ($N=108$, 100 K), 
10 a $N=150$ cluster at 1000 K, 
25 the same cluster in the liquid state at 2200 K,
for a total of 10633 force vectors included in Eq.~(\ref{eq:zf}).

We have used $N_c=32$ additional constraints, 8 of which for
the cohesive energy, the equilibrium lattice spacing $a_\circ$, 
the (unrelaxed) vacancy
formation energy, the (unrelaxed) (111) intrinsic stacking fault 
energy, the (unrelaxed) (111) surface energy, the bulk modulus
and the shear moduli $C_{11}-C_{12}$ and $C_{44}$,
22 to fit the energy and pressure to the 
universal equation of state \cite{rose} at 11 different lattice spacings
($a/a_\circ=$ 0.90, 0.94, 0.97, 1.05, 1.11, 1.20, 1.30, 1.40, 1.50,
1.60, 1.75), and the remaining 2 are related to the invariance
properties of the potential described above.
The weights $W_r$ assigned to the constraints and the cutoff radius
$R_c$ for $\phi(r)$ and $\rho(r)$ have been adjusted
by a trial-and-error procedure, where potentials were generated
by minimizing (\ref{eq:z})
and then run through a test suite including evaluation of
relaxed energies of defects and surfaces, surface relaxations,
thermal expansion and a melting point estimate by 
zero pressure MD simulations.
The final potential, shown in Fig.\ \ref{fig:pot}, 
has $R_c=5.56\,\rm\AA$ (between the 3rd and the 4th neighbor
shell in the fcc crystal), and corresponds to 
$Z=Z_F+Z_C=0.029 + 0.003 = 0.032 \,\rm (eV/\AA)^2$.
$\sqrt{Z_F} \simeq 0.17\,\rm eV/\AA$ is
the root mean square (rms) deviation of force components,
to be compared with the rms force component
in the input data, $0.92\, {\rm eV/\AA}$.
Such error is at least one order of magnitude smaller
than that typical of empirical models \cite{Dave}.
Some properties of the potential are listed in 
Table \ref{tab:calcexp}.
It should be noted that no constraint was imposed on the phonon
frequencies at the zone boundary, so that they are
mostly determined by the force-matching term.
Surface energies are somewhat lower than in experiment,
although they are higher than those predicted by other
potentials of the same class \cite{boston}.
Surface relaxations are very realistic---in
particular, the rather uncommon surface expansion of Al(111)
\cite{alrel111} is obtained---except that the contraction
of Al(110) is not as large as in experiment.
This discrepancy is already present with the first-principle method in 
use \cite{rampi}, and has been simply transmitted to the potential.
The thermal expansion behavior, obtained by
MD at zero pressure, is shown in Fig.\ \ref{fig:thexp}.
The melting temperature, corresponding to the discontinuity,
has been determined with a precision of about 3 K
by achieving solid-liquid coexistence in a
constant enthalpy run for a system with 10752 particles,
following the technique described in \cite{nato}.
Thermal and melting properties are in remarkable agreement with experiment.

In conclusion, this first study shows that the force-matching method 
is a very effective tool to obtain realistic classical potentials
with a high degree of transferability
for systems which the {\em ab initio\/} calculation
technology is capable of treating.
The number of such systems is rapidly increasing
as electronic structure methods are improved
and the computing power increases.
The numerical optimization procedure at the heart of the method
is expected to be well suited to handle easily
the rich and complex analytic forms---including
angular-dependent terms---required for
a realistic modelling of covalent bonds,
and considered difficult to fit so far.
In fact, it could also be used to compare quantitatively
different functional forms, on the basis of their accuracy
in reproducing the {\em ab initio\/} forces.
It would also be of considerable help in the fitting of
alloys, where the number of experimental properties available
is usually rather limited.

We are indebted to Dave Drabold, Cathy Rohrer, Wei Xu and Sang Yang 
for useful discussions and suggestions.
We especially thank Dave Drabold for providing the {\em ab initio} data.
This work has been carried out under the U.S. Department of Energy
Grant No.\ DOE-BES (0) 76ER01198.

\begin{figure}
%\vspace{6cm}
\caption{
The three functions constituting the optimized
glue potential for Al.
}
\label{fig:pot}
\end{figure}

\begin{figure}
%\vspace{6cm}
\caption{
Lattice parameter $a$ as a function of temperature
for our optimized potential (solid line), compared with
experimental data (dotted line).
The jumps indicate the volume change on melting.
In the liquid region, $a$ is defined as $(4\Omega)^{1/3}$
where $\Omega$ is the atomic volume, as in the fcc crystal.
}
\label{fig:thexp}
\end{figure}

\narrowtext
\begin{table}
\caption{
Experimental and calculated (with the optimized potential) values for
equilibrium lattice spacing, cohesive energy,
bulk modulus, elastic constants,
phonon frequencies at the points X, L and K of the Brillouin zone,
vacancy formation and migration energies,
intrinsic (111) stacking fault energy,
surface energy and surface relaxation between
the two outmost layers for the (111), (100) and (110) surfaces,
melting temperature, latent heat and volume change on melting.
All the energies are at $T=0$ and include relaxation effects.
%Melting properties have
%been obtained by achieving solid-liquid coexistence in a MD simulation.
}
\label{tab:calcexp}
%\begin{center}
\begin{tabular}{rcc}
\hline
& exp. & calc. \\
\hline
$a_\circ$ ($\rm\AA$)& 4.032 & 4.032 \\
$E_c$ (eV/atom) & 3.36 & 3.36 \\
$B$ (MBar)      & 0.809 \tablenotemark[1] & 0.809 \\
$C_{11}$ (MBar) & 1.180 \tablenotemark[1] & 1.181 \\
$C_{12}$ (MBar) & 0.624 \tablenotemark[1] & 0.623 \\
$C_{44}$ (MBar) & 0.325 \tablenotemark[1] & 0.367 \\
$\nu_L$ (X)    (THz) & 9.68 \tablenotemark[2] & 9.29 \\
$\nu_T$ (X)    (THz) & 5.81 \tablenotemark[2] & 5.80 \\
$\nu_L$ (L)    (THz) & 9.69 \tablenotemark[2] & 9.51 \\
$\nu_T$ (L)    (THz) & 4.22 \tablenotemark[2] & 4.02 \\
$\nu_L$ (K)    (THz) & 8.67 \tablenotemark[2] & 8.38 \\
$\nu_{T1}$ (K) (THz) & 7.55 \tablenotemark[2] & 7.50 \\
$\nu_{T2}$ (K) (THz) & 5.62 \tablenotemark[2] & 5.34 \\
$E_{\rm vac}^f$ (eV)& 0.66 \tablenotemark[3] & 0.69 \\
$E_{\rm vac}^m$ (eV)& 0.62 \tablenotemark[4] & 0.61 \\
$E_{\rm SF}\,   (\rm meV/\AA^2)$ & 7.5--9.0 \tablenotemark[5] & 6.5 \\
$\gamma_{111}\, (\rm meV/\AA^2)$ & 71--75 \tablenotemark[6] & 54.3 \\
$\gamma_{100}\, (\rm meV/\AA^2)$ & 71--75 \tablenotemark[6] & 58.8 \\
$\gamma_{110}\, (\rm meV/\AA^2)$ & 71--75 \tablenotemark[6] & 64.7 \\
$d_{12}$ (111) (\%) & $+0.9\pm 0.7$ \tablenotemark[7] & $+0.9$ \\
$d_{12}$ (100) (\%) & $-1.2\pm 1.2$ \tablenotemark[8] & $-1.5$ \\
$d_{12}$ (110) (\%) & $-8.5\pm 1.0$ \tablenotemark[9] & $-4.6$ \\
$T_m$ (K)         & 933.6  & $939 \pm 3$ \\
$L_m$ (eV/atom)   & 0.1085 & $0.1054 \pm 0.0003$  \\
$\Delta V_m$ (\%) & 6.5    & $8.4 \pm 0.1$  \\
\hline
\end{tabular}
\tablenotetext[1]{Extrapolated classically to $T=0$ from 
                data in Ref.\ \cite{elacon}.}
\tablenotetext[2]{Frequencies at 80 K from Ref.\ \cite{phonons}.}
\tablenotetext[3]{Ref.\ \cite{vacen}.}
\tablenotetext[4]{Ref.\ \cite{vacmig}.}
\tablenotetext[5]{Ref.\ \cite{sfe}.}
\tablenotetext[6]{Estimates for an ``average'' orientation, 
                  Ref.\ \cite{surfen}.}
\tablenotetext[7]{Ref.\ \cite{alrel111}.}
\tablenotetext[8]{Ref.\ \cite{alrel100}.}
\tablenotetext[9]{Ref.\ \cite{alrel110}.
                  Ref.\ \cite{alrel111} reports $-8.4\pm 0.8$.}
%\end{center}
\end{table}
\widetext

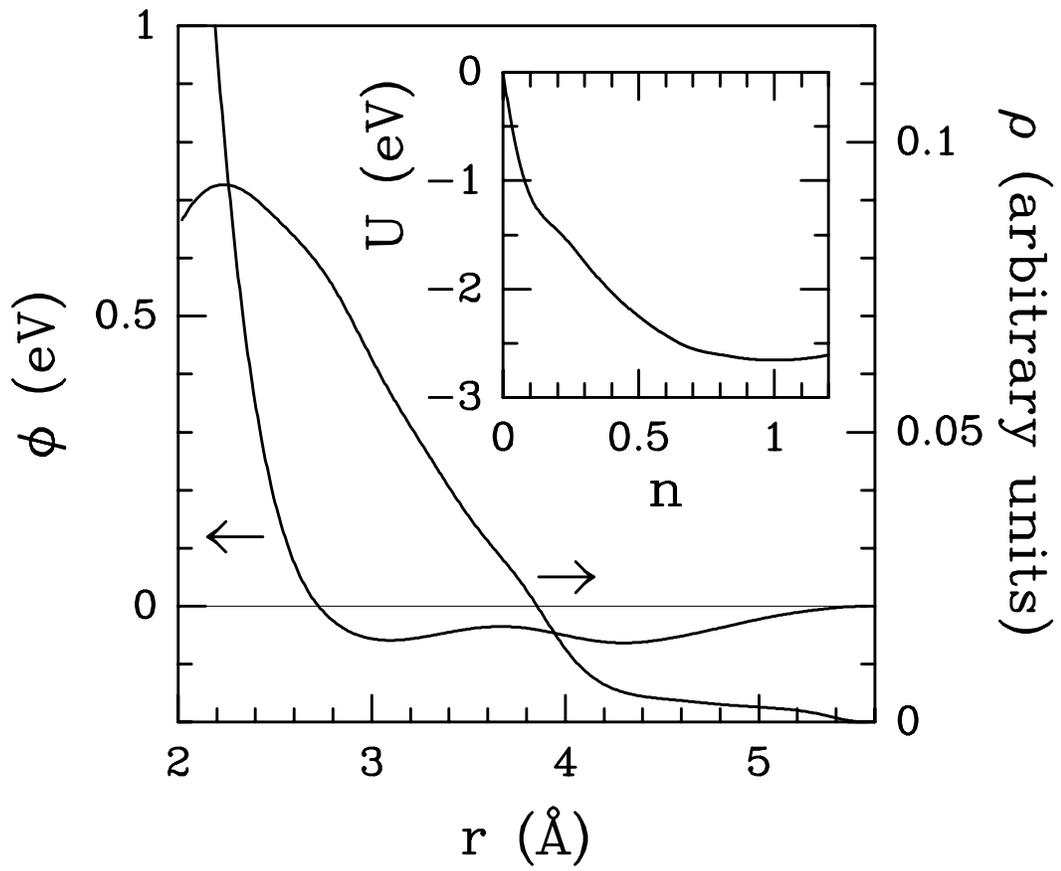
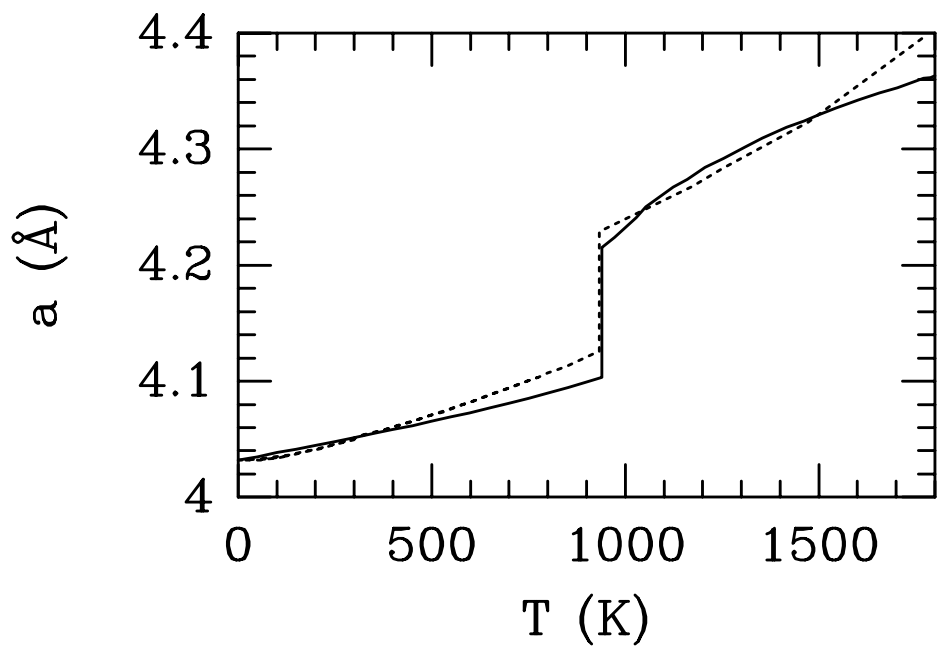
\end{document}